\documentclass[10pt,letterpaper]{article}
\usepackage{opex3}

\usepackage{amsmath,amsfonts,amssymb,graphicx,cite}
\begin{document}

\title{On optical forces in spherical whispering gallery mode resonators}
\author{\normalsize{J. T. Rubin and L. Deych$^*$}}
\address{Department of Physics, Queens College of the City University of New York,  Flushing, NY 11367, USA \& The Graduate Center of CUNY, 365 5th Avenue, New York, NY 10016, USA}
\email{*lev.deych@qc.cuny.edu}

\begin{abstract}
In this paper we discuss the force exerted by the field of an optical cavity on a polarizable dipole. We show that the modification of the cavity modes due to interaction with the dipole significantly alters the properties of the force. In particular, all components of the force are found to be non-conservative, and cannot, therefore, be derived from a potential energy.  We also suggest a simple generalization of the standard formulas for the optical force on the dipole, which reproduces the results of calculations based on the Maxwell stress tensor.
\end{abstract}

\section{Introduction}
Starting with the pioneering works of Ashkin~\cite{Ashkin:70,Ashkin:86}, where trapping and manipulation of objects by optical forces was first demonstrated, there has been an explosion of interest in optical forcing of particles.  Optical tweezers have since been developed into a standard tool used in many applications, while on the fundamental side optical cooling of atoms has opened unique opportunities for exploring various quantum mechanical many-body phenomena.  Recently a great deal of interest has been devoted to the possibility of optical cooling of macroscopic objects such as mechanical nanoresonators~\cite{Anetsberger:2009,Schlesser:2009,SchlesserNatPhys:2009,Groeblacher:2009}, membranes~\cite{Arcizet:2008,Jayich:2008,Wilson:2009} or particles oscillating in optical traps~\cite{Chang:PNAS2010, Romero-Isart:NJP2010, Barker:PRA2010, Romero-Isart:PRA2011,YinPRA:2011,Raizen:NaturePhotonics2011}.  While traditional optical trapping experiments involve freely propagating laser beams, optical microcavities, which confine light in a small volume, have emerged as a candidate for the source of optical forces~\cite{Kippenberg_Science,Schliesser:2010}.  The spatial confinement of light within the small mode volume of the cavity results in an increase of the strength of the  force.  Additionally, a material object interacts with the cavity mode, resulting in a shifted modal frequency and altered field distribution.  As a result, the dynamics of mechanical degrees of freedom become  coupled with those of the field giving rise to so-called dynamical backaction~\cite{Braginsky} responsible for effects such as cavity cooling or heating of mechanical degrees of freedom~\cite{SchlesserNatPhys:2009,Schulze:2010,Barker:PRA2010}.

Understanding cavity optomechanical phenomena depends on correct representation of the optical force exerted by the cavity modes. In the case of free-propagating optical fields (i.e. laser beams), the force on a subwavelength object (dipole) is naturally separated into gradient and scattering components~\cite{Rahmani:2004}.  The gradient component is analogous to the force on a static dipole, which tends to draw a particle into regions of greater field intensity.  It can be presented as the gradient of the electromagnetic energy of the polarized particle and is therefore conservative.  The scattering component results from radiation pressure and is expressed in terms of the momentum flux impinging on the particle per unit time.  This force is non-conservative because it results from the process of irreversible exchange of momentum and energy between the particle and optical field.  Due to its conceptual simplicity and apparent universality, this paradigm has become firmly engrained in the current literature and has been accepted as a framework for calculating optical forces also due to cavity modes.  The effects of spatial confinement are taken into account by using cavity modes to represent electromagnetic field while allowing the resonant frequencies of the cavity to depend on mechanical degrees of freedom.

However, in our recent paper~\cite{Rubin:2011} we pointed out that the validity of the conservative/gradient, non-conservative/scattering paradigm explicitly depends on the assumption that the particle itself does not change sources of the incident field.  While this condition is fulfilled for free-propagating electromagnetic fields, it is violated for fields confined within a cavity.  The position-dependence of the cavity resonance frequencies is just one manifestation of this phenomenon.  In order to elucidate all consequences of the particle-induced modification of the cavity field, we considered the interaction between whispering-gallery-modes of a spherical microresonator and a small dielectric particle.  In this system forces can be calculated by a rigorous analytical approach based on the Maxwell stress tensor.  These calculations show that the force cannot be described within the standard paradigm \cite{Rubin:2011}.  In particular, no vector component of the optical force is conservative, i.e.  derivable as a gradient of potential energy.  Furthermore, the force tangential to the surface of the cavity, which is responsible for the "carousel" effect observed in Ref.~\cite{Arnold:09}, was found to have a contribution proportional to the real part of the particle's polarizability.  This contradicts the assertion of Ref.~\cite{Arnold:09} that the tangential optical force, which is in the direction of the momentum flux of the unmodified by the particle cavity mode, is of strictly "scattering" origin.   In Ref.~\cite{Rubin:2011} these results were obtained using formal \emph{ab initio} approach, which while providing accurate expressions for the force, does not allow for the simple physical interpretation of the results and for finding connections between them and heuristic approaches used in works of other authors. The objective of this paper is to show how the traditional gradient-scattering force approach can be generalized to derive the results of Ref.~\cite{Rubin:2011} without relying on complicated Maxwell stress tensor calculations. It can be conjectured that the "pseudo-gradient" formalism of this paper can be applied even in situations in which rigorous treatment is not possible. In addition, the results presented here allow elucidating limitations of previous heuristic approaches to optical forces.

To achieve this objective in the most efficient way, we organize this paper in the following manner.  We begin by reviewing  the derivations of the gradient and scattering forces paying particular attention to the assumptions involved.  Based on this discussion we propose a psuedo-gradient procedure as a way to extend standard gradient/scattering approach to optical cavities.  We use this idea to determine the force on a particle due to a spherical whispering gallery mode resonator and compare it with the results of stress tensor based calculations.

\section{Optical force on a small polarizable particle}
\subsection{Gradient force: thermodynamic derivation}
We begin by recalling the thermodynamic approach to deriving the electrostatic force on a small dielectric particle.  This approach emphasizes one of the key requirements for the validity of the gradient paradigm. Following the classical textbook by Landau \& Lifshitz~\cite{landau1984electrodynamics} the electrostatic component of the free energy of the polarizable particle is
\begin{equation}\label{eq:total_energy}
u_{tot} = \frac{1}{2} \int{\mathbf{E} \cdot \mathbf{D}}~dV,
\end{equation}
where $\mathbf{D} = \epsilon_0\mathbf{E} + \mathbf{P}$ and the integral runs over the volume of the particle.  $\mathbf{E}$ here refers to the total field in and around the particle, and $\mathbf{P}$ is its polarization. This total energy contains the energy, $u_{ext}$, of the external field, $\mathfrak{E}$, which would have existed in the absence of the particle.  Subtracting $u_{ext}$, the energy required to polarize the particle is
$$
u_{pol}=u_{tot}-\frac{1}{2}\int \epsilon_0|\mathfrak{E}|^2dV
$$
Using this expression one can derive the change in $u_{pol}$ due to a small variation of $\mathfrak{E}$ as
\begin{equation}\label{eq:delta_u}
\delta u_{pol} = -\int \mathbf{P} \cdot \delta \mathfrak{E}dV.
\end{equation}
The force is determined by substituting in $\delta u_{pol}$ the change in the field, $\delta \mathfrak{E}$, due to an infinitesimal translation of the particle, $\delta \mathbf{r}_p$, which is given by: $\delta \mathfrak{E} = (\delta \mathbf{r}_p \cdot \nabla)\mathfrak{E}$.  Equation~\ref{eq:delta_u} then takes the following form:
\begin{equation}
\delta u_{pol} = -\delta \mathbf{r}_p \cdot \int (\mathbf{P} \cdot \nabla) \mathfrak{E}dV.
\end{equation}
Taking into account that this change in energy can be related to the work of the force $\mathbf{F}$ acting on the particle: $\delta u_{pol}=-\mathbf{F}\cdot\delta \mathbf{r}_p$, one obtains expression for the force as
\begin{equation}
\mathbf{F} = \int (\mathbf{P} \cdot \nabla) \mathfrak{E}dV.
\end{equation}
If $\mathfrak{E}$ is approximately constant over the dimensions of the particle (dipole approximation), it can be taken out of the integral, giving
\begin{equation}\label{eq:force_dip_appr}
\mathbf{F} = (\mathbf{p} \cdot \nabla) \mathfrak{E},
\end{equation}
where the dipole moment $\mathbf{p} = \int \mathbf{P}$dV.  Assuming linear polarizability of the particle, $\mathbf{p} = \alpha \mathfrak{E}$, one derives the final gradient expression for the force:
\begin{equation}\label{eq:grad_form}
\mathbf{F} = \frac{1}{2}\alpha_0 \nabla |\mathfrak{E}|^2
\end{equation}
It should be noted, however, that this derivation depends critically on the assumption that a \emph{displacement of the particle does not affect the distribution of the sources of the external field} $\mathfrak{E}$. Without this assumption one would not be able to equate the change of $\mathfrak{E}$ due to small variations of the sources (Eq.~(\ref{eq:delta_u})) with its change due to particle's displacement making the subsequent equations invalid.
\subsection{Gradient force: direct derivation}
\begin{figure}[htbp]
   \centering\includegraphics[width=7cm]{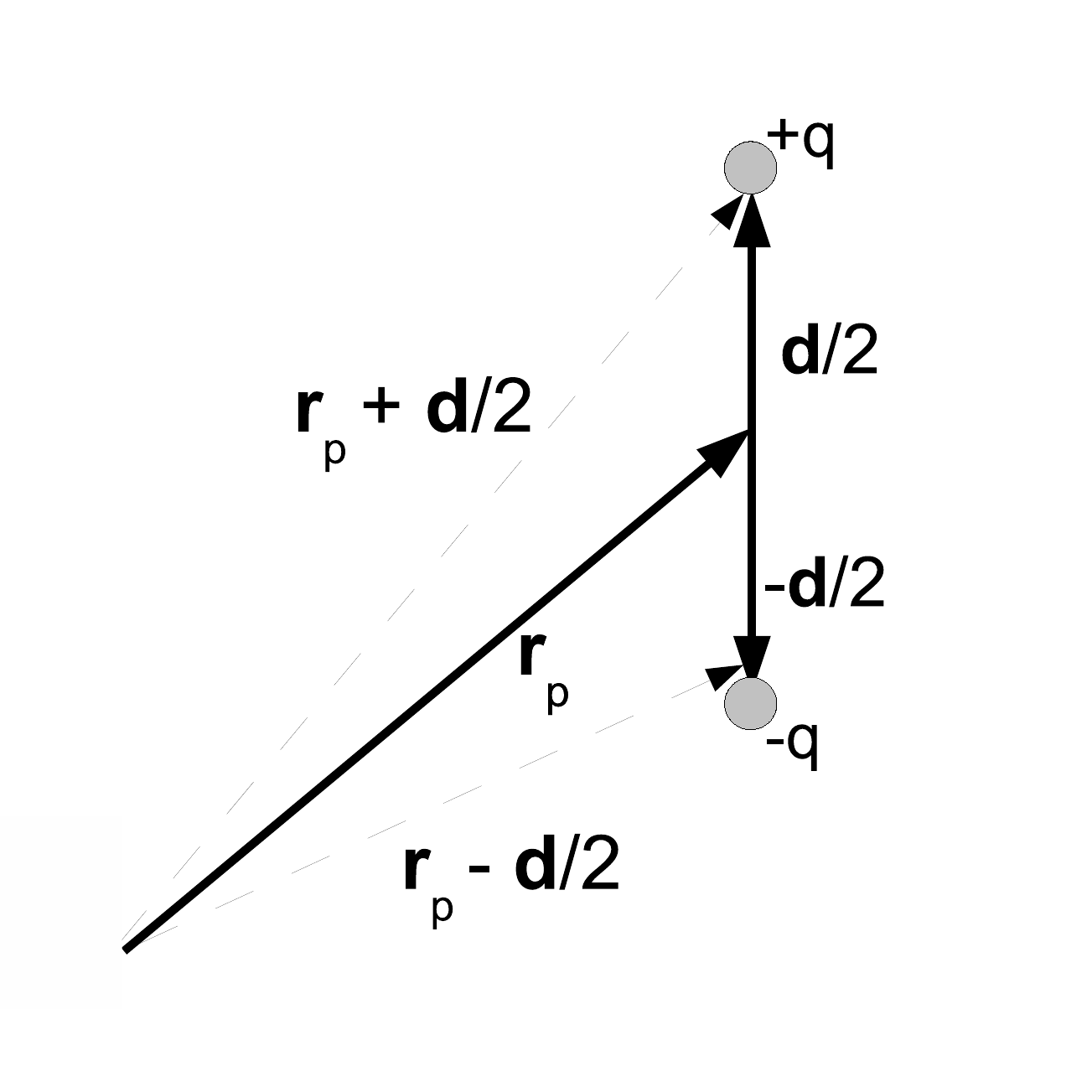}
  \caption{Set up for evaluating the force on a dipole modeled as a system of equal and opposite charges. The distance between the charges will be taken to zero.\label{fig:forces}}
\end{figure}

An alternative derivation, which is also commonly encountered in the textbooks, is based on a model of an electric dipole as a system of equal and opposite charges $\pm q$, separated by some small, ultimately infinitesimal distance $d$. The same approach can be used to describe forces on an induced dipole characterized by polarizability $\alpha$.  The dipole is assumed to be placed in some external field $\mathbf{E}$, so that $\mathbf{p} = \alpha\mathbf{E}$.  We make no demands on $\mathbf{E}$ other than that $\mathbf{p}$ be defined self consistently with it.  In particular, $\mathbf{E}$ may be dependent on the dipole itself.  For example, if the external field $\mathbf{E}$ is due to charges on a conductor, the presence of the dipole will alter the charge distribution and thus $\mathbf{E}$.  We make this explicit by writing $\mathbf{E} = \mathbf{E}(\mathbf{r},\mathbf{r}_p)$, where $\mathbf{r}$ and $\mathbf{r}_p$ are respectively a field point and the position vector of the particle.  The total force is derived by considering the Coulomb forces at each charge comprising the dipole: $\mathbf{F} = q\left[{\mathbf{E}}(\mathbf{r}_p+\mathbf{d}/2,\mathbf{r}_p)-{\mathbf{E}}(\mathbf{r}_p-\mathbf{d}/2,\mathbf{r}_p)\right]$ (see Fig.~\ref{fig:forces}).  By taking the limit $|\mathbf{d}| \rightarrow 0$, keeping $|\mathbf{p}| = q|\mathbf{d}|$ constant, the force derived in this case is:
\begin{equation}\label{eq:force_dipole_direct}
\mathbf{F} = \left[(\mathbf{p} \cdot \nabla_{\mathbf{r}}){\mathbf{E}}\right]_{\mathbf{r} = \mathbf{r}_p}
\end{equation}
where $\nabla_{\mathbf{r}}$ refers to a gradient with respect to field coordinates $\mathbf{r}$.  While this result looks similar to Eq.~(\ref{eq:force_dip_appr}), there is an important difference between  them. The force given by Eq.~(\ref{eq:force_dip_appr}) implies that before taking the spatial derivative of the field with respect to field coordinates $\mathbf{r}$, one sets coordinate of the particle $r_p$ to coincide with $\mathbf{r}$.  On the other hand, Eq.~(\ref{eq:force_dipole_direct}) requires that this procedure is reversed. Physically, it reflects the fact that the electric force on a dipole results from the spatial variation of the electric field across it. The two equations, Eq.~(\ref{eq:force_dip_appr}) and Eq.~(\ref{eq:force_dipole_direct}), produce identical results only if the field exerting the force does not depend upon particle's position.

Equation~(\ref{eq:force_dipole_direct}) can also be transformed into a "gradient" form
\begin{equation}\label{eq:quasigrad}
\mathbf{F} = \frac{1}{2} \alpha \nabla_{\mathbf{r}} |{\mathbf{E}}(\mathbf{r},\mathbf{r}_p)|^2~|_{\mathbf{r} = \mathbf{r}_p}.
\end{equation}
which, however, differs from Eq.~(\ref{eq:grad_form}).  Unlike the latter, Eq.~(\ref{eq:quasigrad}) involves taking the gradient of the function of two variables, and, therefore, the expression $\alpha |{\mathbf{E}}(\mathbf{r},\mathbf{r}_p)|^2/2$ cannot be interpreted as a potential energy unless ${\mathbf{E}}$ is independent of $\mathbf{r}_p$.  Correspondingly, the force calculated according to Eq.~(\ref{eq:grad_form}) does not have to be conservative.


To simplify terminology and notations in the subsequent consideration we will call the operation presented in Eq.~(\ref{eq:quasigrad}) a "pseudo-gradient" and will use notation  $\tilde{\nabla}$ to represent it.
\subsection{Total force on a dipole}
In this subsection we generalize the previous results to the case of a dipole interacting with  a time-dependent harmonic electromagnetic field.  The oscillating dipole moment creates a current density $d\mathbf{p}/dt$, giving rise to a magnetic force.  The total Lorentz force on the particle for incident fields independent of the dipole's position  $\mathfrak{E},\mathfrak{B}=-i/ \omega \nabla \times \mathfrak{E}$ have the standard form $\mathbf{F} = (\mathbf{p} \cdot \nabla)\mathfrak{E} + {d \mathbf{p}}/{dt} \times \mathfrak{B}$. The time averaged expression for this force can be rewritten as \cite{Wong_JOSA_B_2006}:
\begin{equation}\label{eq:gen_force}
\left<\mathbf{F}\right> = \frac{1}{4}\mathcal{R}e[\alpha] {\nabla}|\mathfrak{E}|^2 + \frac{1}{2}\mathcal{I}m[\alpha]\left({\omega} \mathcal{R}e[\mathfrak{E}^{*} \times \mathfrak{B}] + \mathcal{I}m[(\mathfrak{E}^{*} \cdot {\nabla})\mathfrak{E}]\right).
\end{equation}
where $\alpha$ now is a complex valued (with radiative corrections included) polarizability of the dipole. Based upon analysis of the previous sub-section we conjecture that the expression for the force in the case of the field \emph{dependent on the particle's position} $\{\mathbf{E},\mathbf{B}\} = \{\mathbf{E}(\mathbf{r},\mathbf{r}_p),\mathbf{B}(\mathbf{r},\mathbf{r}_p)\}$  can be obtained from Eq.~(\ref{eq:gen_force})  by substituting $\mathfrak{E}, \mathfrak{B}, \nabla$ with $\mathbf{E}, \mathbf{B}, \tilde{\nabla}$ respectively.  {
For a standard dipole particle with radius $R_p$, refractive index $n_p$, and polarizability
$$
\alpha = 4 \pi \epsilon_0(\alpha_0+\frac{2}{3}ik^3\alpha_0^2),
$$
where
$$
\alpha_0 = R_p^3 (n_p^2-1)/(n_p^2+2),
$$
 Equation~(\ref{eq:gen_force}) for the force can be re-written as
\begin{equation}\label{eq:force_final}
\left<\mathbf{F}\right> = -\tilde{\nabla}\left<u\right> + \sigma c\left<\mathbf{g}\right> + \sigma c \frac{\epsilon_0 }{2\omega} \mathcal{I}m[(\mathbf{E}^{*} \cdot \tilde{\nabla})\mathbf{E}],
\end{equation}
where $\left<u\right>$ is the average polarization  energy of the dipole $ \left<u\right> =- \frac{1}{4}\mathcal{R}e[\mathbf{p} \cdot \mathbf{E}]$, $\sigma$ is its scattering cross section $ \sigma = \mathcal{I}m [\alpha] k/\epsilon_0$, and $\left<\mathbf{g}\right>$ is the average momentum density of the field $\left<\mathbf{g}\right> = \frac{1}{2}\mathcal{R}e[\epsilon_0 \mathbf{E} \times \mathbf{B}]$,  and $c$ is the speed of light in vacuum.

In the subsequent sections of the paper we will apply Eq.~(\ref{eq:force_final}) to the case of a small dielectric particle interacting with a whispering gallery mode (WGM) of a spherical resonator. Comparing the obtained expressions for the force with results of Ref.~\cite{Rubin:2011}, where the optical force in this system was calculated using Maxwell stress tensor approach, we will be able to substantiate validity of Eq.~(\ref{eq:force_final}) and shed additional light on physical properties of the optical forces due to cavity-confined electromagnetic field.
\section{Optical force of a WGM resonator}
\subsection{WGMs of a single spherical resonator}
An optical whispering gallery mode is a long living excitation of a spherical resonator, which can be thought of as a ray of light propagating along the equator of the sphere and trapped in it due to total internal reflection. In a spherical coordinate system with  polar axis perpendicular to the plane of the propagation of the mode ($XYZ$ system in Fig.~\ref{fig:coord}), its field is described by a single vector spherical harmonic (VSH) characterized by polar, azimuthal, and radial indexes $l,m,s$ respectively and polarization, TE or TM.  For concreteness we focus only on TE polarized modes with radial index $s=1$ defined as $\mathbf{E} = E_0\mathcal{L}h_l^{(1)}(kr)\mathbf{X}_{l,m}(\theta,\phi)$ (time harmonic factor $e^{-i\omega t}$ is assumed and suppressed).  Here $E_0$ is a normalization factor, $\mathcal{L}$ is the scattering amplitude describing response of the resonator to an incident radiation with frequency $\omega$,  $h_l^{(1)}(kr)$ is the spherical Hankel function of the first kind, and $k$ is the magnitude of the wavevector $k = \omega/c$. In the close vicinity of a chosen resonant frequency the scattering amplitude can be approximated as
\begin{equation}
 \mathcal{L} = -i\Gamma_l^{(0)}/(\omega-\omega_l^{(0)}+i\Gamma_l^{(0)})
\end{equation}
with $\omega_l^{(0)}$ and $\gamma_l^{(0)}$ being respectively the resonant frequency and decay rate of the mode.  The angular portion of the VSH is defined as $\mathbf{X}_{l,m} = \mathbf{L}Y_{l,m}(\theta,\phi)/\sqrt{l(l+1)}$, where $\mathbf{L}$ is the dimensionless angular momentum operator $\mathbf{L} = -i \mathbf{r} \times \nabla$, and $Y_{l,m}(\theta,\phi)$ is the scalar spherical harmonic.  We focus here on so called fundamental modes with $m = l$.  WGMs with long life-time are characterized  by $l \gg 1$ and in expressions that follow we neglect terms of order $1/l$ compared to those or order unity.
 \begin{figure}[htbp]
 \centering\includegraphics[width=7cm]{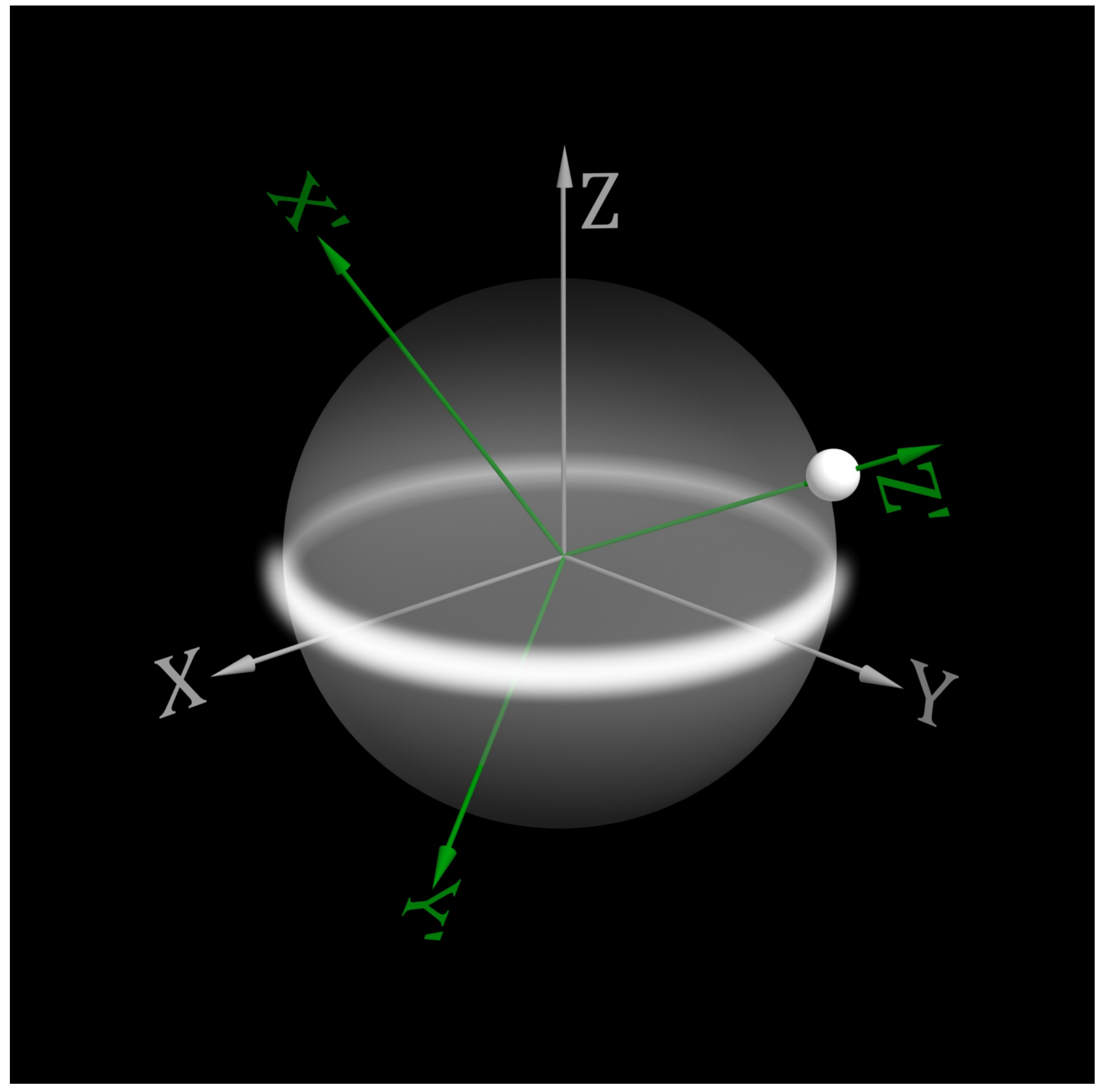}
    \caption{Coordinate systems used for evaluation of optical forces together with schematic presentation of the resonator, WGM, and the dipole.  The  axis always connects the center of the resonator and the point of observation. When using this coordinate system to calculate the forces, this axis passes through the center of the particle. \label{fig:coord}}
\end{figure}

For calculation of optical forces due to this WGM it is also convenient to find an expression for its field in a coordinate system with polar axis connecting the center of the sphere and the point of observation ($X^\prime Y^\prime Z^\prime$ system in Fig.~\ref{fig:coord}.) Such an expression can be obtained using rotational properties of VSH~\cite{Mishchenko_book2002} expressed as
\begin{equation}\label{eq:rot}
\mathbf{X}_{l,m}(\theta^{\prime},\phi^{\prime}) = \sum_{m^{\prime}=-l}^l D_{m^{\prime},m}^{(l)}(\alpha,\beta,\gamma)\mathbf{X}_{l,m^{\prime}}(\theta,\phi)
\end{equation}
where $D_{m^{\prime},m}^{(l)}(\alpha,\beta,\gamma)$ is the Wigner D function and $\alpha,\beta,\gamma$ are the Euler angles specifying the rotation from the unprimed to primed coordinate system.  The D-functions are defined by
$$
D_{m^{\prime},m}^{(l)}(\alpha,\beta,\gamma) = e^{-i (m^{\prime} \alpha + m\gamma)}d_{m^{\prime},m}^{(l)}(\beta),
$$
where the function $d_{m,l}^{(l)}(\beta)$ for $m = l$ can be written as
\begin{equation}
d_{m,l}^{(l)}(\beta) = \sqrt{\frac{(2l)!}{(l+m)!(l-m)!}}\left[\cos{\frac{\beta}{2}} \sin{\frac{\beta}{2}}\right]^{l}\left[\cot(\frac{\beta}{2})\right]^m
\end{equation}
In order to find an expression for the WGM in the primed system one has to apply inverse transformation $D_{m^{\prime},m}^{(l)}(-\gamma,-\beta,-\alpha) = [D_{m,m^{\prime}}^{(l)}(\alpha,\beta,\gamma)]^{*}$ to the VSH defined in the unprimed coordinate system.  Rotation by the angle $\gamma$ is equivalent to shifting the $\phi^{\prime}$ coordinate, so we  only consider transformations whith $\gamma = 0$.  Applying this transformation to a WGM with orbital number $l=L$ we find its representation in the rotated system in the following form:
\begin{equation}\label{eq:fieldrot}
{\mathbf{E}}^{\prime} = E_0h_L^{(1)}\sum_{m^{\prime}}a_{m^{\prime}}\mathbf{X}_{L,m^{\prime}}
\end{equation}
where
\begin{equation}\label{eq:am}
a_{m} =  \mathcal{L} D_{m,L}^{(L)}(0,-\beta,-\alpha) = \frac{-ie^{i L \alpha}d_{m,L}^{(L)}(-\beta)}{y_0 +i}
\end{equation}
Here we introduced dimensionless variable $y_0 = (\omega - \omega_L^{(0)})/\Gamma_L^{(0)}$ representing the relative detuning of the external frequency from the resonance of the WGM with respect to its width. At any point on the polar $Z^{\prime}$ axis ($\theta^{\prime}=0$), all $\mathbf{X}_{L,m}$ vanish except for $\mathbf{X}_{L,\pm1}$.   Assuming $L \gg 1$, and dispensing with the prime on $\mathbf{E}^\prime$, the field can be written explicitly as:
\begin{equation}\label{eq:e0field}
\mathbf{E} = E_0h_L^{(1)}(kr)\sqrt{\frac{L}{4 \pi}}\left(a_{-1} \hat{\mathbf{\xi}}_{+} + a_{1} \hat{\mathbf{\xi}}_{-}\right)
\end{equation}
where $\xi_{\pm} = (i\hat{\mathbf{\theta}} \pm \hat{\mathbf{\phi}})/\sqrt{2}$ and $\hat{\mathbf{\theta}}$ and $\hat{\mathbf{\phi}}$ are the spherical coordinate unit vectors referred to the global, unprimed system.  The magnetic field $\mathbf{H} = \mathbf{B}/\mu_0 = -i/ \omega\mu_0 \nabla \times \mathbf{E}$ is given by:

\begin{equation}\label{eq:h0field}
\mathbf{H} = E_0 \sqrt{\frac{\epsilon_0}{\mu_0}} \sqrt{\frac{L}{4 \pi}}\left[-i \sqrt{2} L \frac{h_L^{(1)}(kr)}{kr}a_0 \hat{r} + \frac{[kr h_L^{(1)}(kr)]^{\prime}}{kr} \left(a_1 \hat{\mathbf{\xi}}_{+} - a_{-1} \hat{\mathbf{\xi}}_{-}\right)\right]
\end{equation}
where the prime denotes differentiation with respect to argument $kr$.  We can express the field at any point in space in the form of Eq.~(\ref{eq:e0field}) and Eq.~(\ref{eq:h0field}) by changing the Euler angles appearing in $a_{m}$.  It is important to note that angles $\alpha$ and $\beta$ correspond to the respective angular coordinates $\phi$ and $\theta$ of the $Z^{\prime}$ axis as viewed from the unprimed coordinate system.
\subsection{Calculation of the force neglecting particle-resonator coupling}
In order to elucidate the effects of the particle-induced modification of WGMs on the properties of the optical force, we first compute this force with this modification neglected. In this case, assuming that a particle with coordinates $\mathbf{r}_p =(r_p,\theta_p,\phi_p)$ as defined in the $XYZ$ coordinate system, lies on the $Z^\prime$ axis of the $X^\prime Y^\prime Z^\prime$ system, we can substitute the results of the previous sub-section, Eq.~(\ref{eq:e0field}) and (\ref{eq:h0field}), into Eq.~(\ref{eq:force_final}) for the force, while replacing the operator of pseudo-gradient with the regular gradient.
In the limit $L \gg 1,m$ function $d_{m,L}^{(L)}(-\beta)$ can be approximated as
\begin{equation}
d_{m,L}^{(L)}(-\theta_p) \approx \frac{1}{(\pi L)^{1/4}}{e^{-\frac{L}{2}\left({\theta_p-\frac{\pi}{2}}\right)^2}}\left[\cot\left(-\frac{\theta_p}{2}\right)\right]^m
\end{equation}
where we took into account that $\beta\equiv \theta_p$. Consequently, the field coefficients $a_m$ for $m=\pm 1$ can be presented in terms of the $m=0$ coefficient $a_0$ as $a_m = a_0 [\cot{\theta_p/2}]^m$. It is clear, therefore, that products of the form $a_{m}^{*} a_{m^\prime}$ in this case are purely real, so that any terms in the force proportional to $\mathcal{I}m [a_{m}^{*} a_{m^\prime}]$ vanish.  Since the field strength decays quickly when the particle moves out of the equatorial plane $\theta_p = \pi/2$, we only consider  $\bar{\theta}=\theta_p-\pi/2\ll 1$, and expand $d_{m,L}^{(L)}(-\theta_p)$  in terms of $\bar{\theta}$. However, since for $L \gg m$, $L\bar{\theta}\gg \bar{\theta}$ we shall only expand the terms, which do not contain $L$. Keeping  linear terms in the expansion of  $[\cot{\theta_p/2}]^m$, and taking into account that for $L\gg 1$ the imaginary part of the Hankel function $h_L^{(1)}(\rho)$ is much greater than its real part in the region $\rho < L$, we obtain the gradient portion of the optical force $\mathbf{F}^{(g)}= \nabla\left<\mathfrak{u}\right>$:
\begin{equation}\label{eq:grad_force_unmod}
\mathbf{F}^{(g)}  = \mathfrak{F}\frac{1}{y_0^2+1}\left([n_L(kr_p)]^{\prime}\hat{r} - \frac{L}{kr_p}n_L(kr_p)\bar{\theta}\hat{\theta}\right)
\end{equation}
where
\begin{equation}\label{eq:force_amplitude}
\mathfrak{F} = \frac{1}{4\pi^{3/2}}\mathcal{R}e[\alpha] |E_0|^2 k\sqrt{L}n_L(kr_p)e^{-L\bar{\theta}^2}
\end{equation}
and $n_L(kr_p)$ is the spherical Neumann function.

The leading term in $\sigma c \frac{\epsilon_0 }{2\omega} \mathcal{I}m[(\mathbf{E}^{*} \cdot \tilde{\nabla})\mathbf{E}]$ is of the second order in $\bar{\theta}$ and can be neglected compared to the scattering force,  $\mathbf{F}^{(s)}=\sigma c \left<\mathbf{g}\right>$, which takes the form:
\begin{equation}\label{eq:scat_force_unmod}
\mathbf{F}^{(s)}= \mathfrak{F} \frac{p}{y_0^2 +1 }\frac{L}{kr_p}n_L(kr_p)\hat{\phi}
\end{equation}
where $p=2k^3\alpha_0/3\ll 1$.
Thus, in this approximation, the gradient force draws the particle toward the resonator radially, and maintains it in the equatorial plane via its $\hat{\theta}$ component. It is important to note that its dependence on the particle's position follows the behavior of the Neumann function $n_L(kr)$, which is almost exponential in the considered range of parameters. At the same time, the azimuthal component of the force is of purely scattering nature and proportional to the moment density, which can be presented in the simple form $\left<\mathbf{g}\right> = (L/kr)u_{em}\hat{\phi}$, where $u_{em} = \epsilon_0\left<|\mathfrak{E}|^2\right>$ is the electromagnetic energy density (the use of $\mathfrak{E}$ symbol for the field emphasizes that it is not affected by the presence of the particle).
\subsection{Effect of the particle-induced modification of the WGM on the optical force}
\subsubsection{Modification of the WGM by the particle}
The problem of determination of the electromagnetic field of the coupled resonator-dipole system is analytically tractable and was solved in Ref.~\cite{DeychPRA:2009,Rubin:2010}. The dipole is modeled as a small sphere with radius $R_p$, where $kR_p \ll 1$, and refractive index $n_p$, and the field is found in the form of a general expansion in terms VSHs with all $l,m$ and polarizations.  The particle is found to modify the WGM in two significant ways.  First, it creates an additional resonance at frequency $\omega_p = \omega_L^{(0)} + \delta \omega_L$ with width $\Gamma_p = \Gamma_L^{(0)} + \delta \Gamma_L$ in addition to the  $\omega_L^{(0)}$ resonance of a single sphere, with $\delta \omega_L$ and $\delta \Gamma_L$ depending only on $r_p$.  Second, the steady state of the resonator field associated with the $\omega_p$ resonance is significantly modified compared to the field distribution of the initial WGM. Initially isotropic field turns into a highly directional distribution oriented predominantly toward the particle.  Thus, a displacement of the particle in the $\hat{\phi}$ direction causes the resonator's field to move with the particle (see for details Ref.~\cite{DeychPRA:2009,Rubin:2010}).  In addition, the interaction with the particle  excites in the resonator WGMs with different $l$ and polarization.  However, these contributions are small, and can be neglected. In this approximation, the scattered field of the resonator can again be presented in the form of Eq.~(\ref{eq:fieldrot}), but with expansion coefficients, which are no longer given by Eq.~(\ref{eq:am}). They have the following form
\begin{eqnarray}\label{eq:ammod}
 a_{m}&=&{-ie^{i L \phi_p}d_{m,l}^{(l)}(-\theta_p)} \times \left\{ \begin{array}{lcc}
 \displaystyle{[y_0 + i]^{-1}} & {m \ne \pm1}  \\
 \displaystyle{\frac{\Gamma_L^{(0)}}{\Gamma_p}[y + i]^{-1}} & m = \pm1 &
 \end{array}\right\}
\end{eqnarray}
where $y = (\omega - \omega_p)/\Gamma_p$.  The frequency shift and additional broadening of the resonance are~\cite{DeychPRA:2009,Rubin:2010}:
\begin{equation}\label{eq:delt}
\delta \omega_L = -\frac{\mathcal{R}e[{\alpha}]k^3}{6 \pi \epsilon_0}\Gamma_L^{(0)}[V_{L,1}(kr_p)]^2;\hskip 2pt \delta \Gamma_{L} = {p}|\delta \omega_L|
\end{equation}
where $V_{L,1}(kr_p)$ is the VSH translation coefficient~\cite{Mishchenko_book2002}, which arises when the field scattered by one sphere is expressed in terms of VSH centered about the other, and is given by
$$
{V}_{L,m}(kr_p) = i(-1)^{L+1}\frac{\sqrt{3}}{2}m\sqrt{2L+1}h_L(kr_p)^{(1)}  \approx (-1)^{L}\sqrt{3L/2}n_L(kr_p).
$$
\subsubsection{Calculation of the force with particle-modified field}
In this part of the paper we assess the effects of the particle-induced shift of the resonance frequency and of the changes in the spatial configuration of the field of the resonator on the optical forces exerted by it. To this end we shall analyze the expressions for the force obtained by evaluating Eq.~(\ref{eq:force_final}) with the field at the location of the particle given by  Eq.~(\ref{eq:fieldrot}). The role of the pseudo-gradient operator in this equation is to distinguish between field coordinates $\mathbf{r}$ and particle coordinates $\mathbf{r_p}$ even though we calculate the force at the point $\mathbf{r}=\mathbf{r_p}$. Taking into account that dependence on $\mathbf{r_p}$ is only contained in the expansion coefficients $a_m$, this procedure becomes rather trivial: one needs to find the gradient of all respective expressions treating these coefficients as constants, and after that equate $\mathbf{r}=\mathbf{r_p}$. Calculating the required gradients we obtain for the pseudo-gradient component of the force: $\mathbf{F}^{(pg)}\equiv\tilde{\nabla} \langle u\rangle$
\begin{eqnarray}\label{eq:quasigradient_force}
F^{(pg)}_r &=&\frac{1}{4}\epsilon_0 |E_0|^2L\alpha_0\left(|a_{-1}|^2+|a_{1}|^2\right)\left.\frac{d |h_L^{(1)}(kr)|^2}{d r}\right|_{r=r_p}\\
F^{(pg)}_\theta& =& \frac{1}{4}\epsilon_0|E_0|^2|h_L^{(1)}(kr_p)|^2\frac{L^2\alpha_0}{r_p \sin{\theta}} \mathcal{R}e \left[  a_{1}(a_0^{*} - a_2^{*}) - a_{-1}(a_0^{*} - a_{-2}^{*})\right]\\
F^{(pg)}_\phi  &=& -\frac{1}{4}\epsilon_0|E_0|^2|h_L^{(1)}(kr_p)|^2\frac{L^2\alpha_0}{r_p} \mathcal{I}m \left[  a_{1}(a_0^{*} + a_2^{*}) + a_{-1}(a_0^{*} + a_{-2}^{*})\right]
\end{eqnarray}
The scattering component of the force, $\mathbf{F}^{(s)}$, takes in this case the form of
\begin{equation}\label{eq:newsc_force}
\mathbf{F}^{(s)}  = {\epsilon_0}|E_0|^2\frac{L^2\alpha_0^2k^3}{3 r}|h_L^{(1)}(kr)|^2 \left\{\hat{\theta} \mathcal{I}m\left[(a_{0}^{*}(a_{-1} - a_{1})\right] + \hat{\phi}\mathcal{R}e\left[(a_{0}^{*}(a_1 + a_{-1})\right]\right\}
\end{equation}
A contribution from the remaining term in Eq.~(\ref{eq:force_final}), which is proportional to $\mathcal{I}m (\mathbf{E}^{*} \cdot \tilde{\nabla})\mathbf{E}$ remains negligible and will not be considered any further.

One can see that Eq.~(\ref{eq:quasigradient_force}) and Eq.~(\ref{eq:newsc_force}) differ significantly from the respective Eq.~(\ref{eq:grad_force_unmod}) and Eq.~(\ref{eq:scat_force_unmod}) obtained under the assumption of the unmodified WGM. Further analysis of the obtained expression will be performed in two steps. Since it is often assumed that the main effect of the particle on the cavity mode consists in changing the resonance frequency, we first separate this effect. To achieve this we allow the particle to shift and broaden the resonance according to Eq.~(\ref{eq:delt}), but will assume that the field coefficients are given by the unmodified Eq.~(\ref{eq:am}) with replacement of $(y_0+i)^{-1}$ by $(\Gamma_p/\Gamma_L^{(0)})(y+i)^{-1}$. In this case the particle can modify the amplitude of the resonator's field, but does not change its spatial distribution.   Then, one immediately sees that in Eq.~(\ref{eq:quasigradient_force}), $F^{(pg)}_\phi$, which contains terms proportional to $\mathcal{I}m[a_ma_{m^\prime}]$ with $m\ne m^\prime$ vanishes. For the same reason the $\theta$-component of the scattering force also vanishes.

In order to present the radial component in the form easily related to expressions used in papers of other authors such as those of Ref.~\cite{Chang:PNAS2010,Romero-Isart:PRA2011,Hu:2010}, we introduce  the power radiated by those modes of the resonator that interact with the particle as,
$$
P = |E_0|^2\frac{\epsilon_0 c^3}{2\omega^2} \left(|a_{1}|^2+|a_{-1}|^2\right)=|E_0|^2\frac{\epsilon_0c^{3}}{\omega^2}[d_{0,L}^{L}]^2\left(\frac{\Gamma_L^{(0)}}{\Gamma_p}\right)^2\frac{1}{y^2+1},
$$
and the number of photons in them as
$$
N = \frac{P[2\Gamma_L^{(0)}]^{-1}}{\hbar\omega}.
$$
The expression for the radial component of the force can be written down now as
\begin{equation}\label{eq:frmod}
F_r = -N\hbar\frac{d \delta \omega_L}{dr_p},
\end{equation}
which should be contrasted with the expression obtained from the usual derivative of interaction energy
\begin{equation}\label{eq:dupol}
\frac{du_{pol}}{d r_p} = N\hbar\frac{d \delta \omega_L}{dr_p} + \frac{d N}{d r_p}\hbar \delta \omega_L
\end{equation}
The last term in this expression is spurious as was first noticed in Ref.~\cite{Hu:2010}, where it was shown by direct numerical simulations that the force on a dipole in Fabry-Perot resonator must be given by Eq.~(\ref{eq:frmod}) rather than by Eq.~(\ref{eq:dupol}).

This result is a clear demonstration of the fact that the polarization energy of the particle cannot be considered a true potential energy even if one neglects the spatial modification of the cavity mode due its interaction with the particle. However, taking this modification into account results even in more drastic changes in the optical force yielding a non-zero azimuthal component of the pseudo-gradient force and  a non-zero polar component of the scattering force. The expression for the total force $\mathbf{F}=\mathbf{F}^{(pg)}+\mathbf{F}^{(s)}$ in this case is found by using the correct set of the field coefficients as defined by Eq.~(\ref{eq:ammod}). The radial component of the force does not change from Eq.~(\ref{eq:frmod}), while its polar and azimuthal components now can be presented as
\begin{equation}\label{eq:Ftheta_full}
F_\theta = 2\hbar N\frac{L\Gamma_p}{r_p\Gamma_L^{(0)}(y_0^2+1)(y^2+1)}\bar{\theta}\left[{\delta \omega_L(1+yy_0) + \delta \Gamma_L(y_0 - y)}\right]
\end{equation}
\begin{equation}\label{eq:Fphi_full}
F_\phi =2\hbar N\frac{L\Gamma_p}{r_p\Gamma_L^{(0)}(y_0^2+1)(y^2+1)}\left[{\delta \omega_L(y_0-y) + \delta \Gamma_L(1 + y_0y)}\right]
\end{equation}
Terms proportional to $\delta \omega_L$ come from the psuedo-gradient force, while $\delta \Gamma_L$ terms come from the scattering force. These expressions demonstrate significant deviation of the force from both completely unmodified and frequency-only modified WGM approximations. First, let us note that the radial dependence of the force is determined by the factor  $(y^2+1)^{-1}$ in addition to the Neumann  function in $\delta\omega_L$ and $\delta\Gamma_L$.  The role of this factor  can be seen as follows.  The condition $y=0$ is satisfied for some $r_p = r_0$ at which driving frequency $\omega$ coincides with the particle-induced resonance.  If one linearizes   $y(r_p)$ about this point as $y = (r_p-r_0)y^{\prime}(r_0)$, where
$$
y^{\prime}(r_0) =- \frac{1}{\Gamma_p}\frac{d \delta \omega_L}{dr_p}\left(1 + yp\right)
$$
If $p$ is sufficiently small, then in the region where $y^{\prime}(r_0)$ can be considered constant, the factor $d \delta \omega_L/dr_p$ in $F_r$ is also constant.  Therefore the spatial profile of the force  has \emph{Lorentzian shape} peaked at $r_p = r_0$ with width $1/y^{\prime}$.  Let us recall that in the unmodified WGM approximation the magnitude of the force monotonically (essentially exponentially) decreases with $r_p$.

The azimuthal force $F_{\phi}$ is no longer solely due to the scattering contribution. Two different limits are of interest based upon choice of the external driving frequency $\omega$.  In the limit $\omega \rightarrow \omega_L^{(0)}$, $F_{\phi} \propto y\delta \omega_L + \delta \Gamma_L$.  The magnitude of the pseudo-gradient term exceeds that of the scattering term unless $y \ll p$.  This can only happen for very small values of $\delta \omega_L$, when the magnitude of the force is also very small.  When $y \gg p$, the pseudo-gradient contribution exceeds the scattering force, and the tangential component can be written as $F_{\phi} = (F_\phi^{(0)}\Gamma_L^{(0)})/(yp\Gamma_p)$, where $F_{\phi}^{(0)}$ is the scattering force in the unmodified WGM approximation as given by Eq.~(\ref{eq:scat_force_unmod}). If $y$ satisfies $1 \ll y < \Gamma_L^{(0)}/(p\Gamma_p)$, the tangential force exceeds $F_{\phi}^{(0)}$.  When $\omega-\omega_L^{(0)} \gg \Gamma_L^{(0)}$ the scattering contribution to $F_{\phi}$ becomes negligible as well.  In this case the force can be written $F_{\phi} = F_{\phi}^{(0)}\Gamma_L^{(0)}/p\Gamma_p$, which also exceeds the magnitude that would have been obtained in the unmodified WGM approximation. These results show that the force propelling the particle in the experiments like the one of Ref.~\cite{Arnold:09} is not necessarily of scattering origin and might have a pseudo-gradient contribution.  The two can be distinguished by their dependence on $\alpha_0$: while the pseudo-gradient force is linear in this parameter, the scattering force is quadratic.

Even in the range of parameters where the pseudo-gradient contribution to the azimuthal component of the force dominates, it remains non-conservative since it imparts net kinetic energy to the particle moving along a closed orbit around the resonator.  This occurs because there is a field gradient which pushes the particle in the $\hat{\phi}$ direction.  When the particle moves to a new point, the field re-adjusts so that there is again a field gradient in the $\hat{\phi}$ direction.  Implicit in this analysis is the assumption that the particle moves slowly enough to consider the field always remaining in the quasi-steady state.  Velocity dependent effects can become significant when the time scale of particle motion is comparable to the relaxation time of the resonator, $1/\Gamma_p$.

The relative magnitudes of the force components can be analyzed by comparing $(L/r_p)\delta \omega_L$ to $d \delta \omega_L/d r_p$.  From the asymptotic expansions for the spherical Nuemann functions and their derivatives in the region $L \gg 1$, $kr < L$, we have $n_L/n_L^{\prime} \approx -cosh(a)$ \cite{AbStegun}  where prime denotes differentiation with respect to argument and $a$ is defined by $kr_p = (L+1/2)sech(a)$.  Since resonances are in the region $kr_p \approx L/n$, where $n$ is the refractive index of the resonator, we have $n_L/n_L^{\prime} \approx -n$ and thus $(d \delta \omega_L/d r_p)/(L \delta \omega_L/r_p) \approx -2n k r_p/L \approx 2$.  Thus, assuming we are near a particle induced resonance so that $0 \le |y| \le O(1)$, the relative magnitude of the forces will be determined by the factor $(y_0^2 + 1)^{-1}$ in $F_{\phi}$ and $F_{\theta}$.  If the system is driven at a frequency close to the ideal Mie frequency, so that $y_0$ is of order unity, then the radial and azimuthal forces will be of comparable magnitude, while the polar force will be smaller by a factor $\bar{\theta}$.  If $y_0 \gg 1$ on the other hand, then the azimuthal force will be smaller than the radial force by a factor $1/y_0$, while the polar force is smaller by a factor $\bar{\theta}y/y_0$.

It can also be seen that the scattering contribution to the azimuthal and polar forces is in general smaller than the pseudo-gradient contribution due to the fact that $\delta \Gamma_L / \delta \omega_L = p \ll 1$ (where $p=2k^3\alpha_0/3$).  In the limit where both the driving frequency and particle induced resonance frequencies are very close to the ideal Mie resonance, so that $y,y_0 \rightarrow 0$, the scattering contribution to $F_{\phi}$ becomes appreciable, while it vanishes in $F_{\theta}$.  This is to be expected given that $y,y_0 \rightarrow 0$ is the limit where the particle induced modification of the cavity mode becomes vanishingly small, and accordingly the behavior of the forces approaches that of their unmodified forms of Eq.~(\ref{eq:grad_force_unmod}) and Eq.~(\ref{eq:scat_force_unmod}).  This is likely the regime encountered in the experiments of Ref.~\cite{Arnold:09}.

The results of the calculation of the force within the pseudo-gradient approach can be compared with  calculations carried out by integrating the Maxwell stress tensor over a surface of the particle, which is assumed to have a small, but finite size.  For a field represented by a VSH expansion, the stress tensor integral over a spherical region can be performed analytically and the force given in terms of the VSH expansion coefficients~\cite{Chen_PRE_2009}.  For the present case these coefficients are given in Ref. \cite{Rubin:2010} while full details of the calculations of the force can be found in Ref.~\cite{Rubin:2011}.   In the large $L$ limit, the forces obtained agree exactly with those calculated from the pseudo-gradient approach validating the latter.

It is interesting to note that the large $L$ limit of the stress tensor calculations is necessary to maintain consistency  with the assumed point-like nature of the particle in the pseudo-gradient approach. To see this, note that for a given resonator of radius $R$ and refractive index $n$, the lowest order approximation to the resonant frequency is the geometric optical condition $nkR \approx L$.  At the same time, a point dipole is defined by the the limit $R_p \rightarrow 0$ with electromagnetic size parameter $\rho = k R_p$ kept constant.  Combining these two conditions we have $nR\rho=LR_p$, which implies that taking $R_p \rightarrow 0$ requires that $L \rightarrow \infty$.
\section{Conclusion}
We have presented here a generalization of the theory of  optical forces on a dipole for the case when it interacts with the electromagnetic field of an optical cavity.  The traditional gradient/scattering paradigm is shown to be invalid when the dipole can modify the source of the field.  In particular, all vector components of the force are found to be non-conservative and, consequently, no component can be derived from a gradient of electromagnetic polarization energy.  We have further shown that, when the particle-induced modification of the resonator field is taken into account, the force in the direction of the energy and momentum flux of the wave cannot be interpreted as a 'scattering' force.  In place of the gradient/scattering paradigm we have proposed a pseudo-gradient framework which is conceptually simpler and computationally more efficient than the exact Maxwell stress tensor approach. Using an example of a small dielectric particle interacting with whispering-gallery-modes of a spherical resonator we demonstrated that the suggested pseudo-gradient formalism reproduces all results of the  calculations based on the Maxwell stress tensor.

The results of this work have important implications for the quantum theory of optomechanical interaction, which is commonly based on the assumed potential nature of the gradient force.  These results are also of importance for proposed optofluidic sensors which rely on a tangential force to drive the particle in orbit around the resonator.

\end{document}